\documentclass[prl,superscriptaddress,twocolumn,showkeys,amsmath,%
preprintnumbers]{revtex4}
\usepackage{bm} 
\usepackage{graphicx}
\newcommand{\beq}{\begin{equation}}
\newcommand{\eeq}{\end{equation}}
\newcommand{\be}{\begin{eqnarray}}
\newcommand{\ee}{\end{eqnarray}}
\usepackage{epstopdf}
\usepackage{epsfig}
\newcommand \Pomeron {I\!\!P}
\newcommand \jp {$J/\psi \,$}
\begin{document}
\author{L.~Frankfurt}
\affiliation{School of Physics and Astronomy, Tel Aviv University, 
Tel Aviv, Israel}
\author{M.~Strikman}
\affiliation{Department of Physics, Pennsylvania State University,
University Park, PA 16802, USA}
\author{M. ~Zhalov}
\affiliation{St Petersburg Nuclear Physics Institute, Gatchina, Russia}
\title{Tracking fast small color dipoles through 
strong gluon fields at the LHC
}

\begin{abstract}
We argue that  the process 
$\gamma +A \to J/\psi + "{\rm \,  gap"} + X$ at large momentum transfer
$q^2$ provides a quick and  effective way to test onset of a novel perturbative QCD regime of 
strong absorption for the interaction of small 
dipoles at the collider energies. We find that already the first heavy ion run at the LHC
 will allow to study this reaction with  sufficient statistics via  ultraperipheral  collisions 
hence probing the interaction of $q\bar{q}$  dipoles of sizes $\sim {\rm 0.2 fm}$ with nuclear media 
down to $x\sim 10^{-5}$.

\end{abstract}


\maketitle

Soon after $J/\psi$ was discovered, the $J/\psi$ photoproduction experiments on nuclear targets
have established 
that  nuclei are practically transparent to the $J/\psi$'s produced at  photon energies in the 
range $\sim \rm{20 \,GeV}\div 120 GeV$ \cite{Anderson:1976hi,Sokoloff:1986bu}.
The absorptive cross 
section, $\sigma_{abs}^{J/\psi N}$, was found to be close to $\sim 4 \, {\rm mb}$ 
that
is much smaller than the cross section of interaction of ordinary mesons $\sim 25 \, {\rm mb}$. The observed transparency is natural within the Low-Nussinov model of two gluon exchange where the
 cross section of hadron interaction with a  small color singlet dipole quark-antiquark 
configuration in the photon wave function 
is proportional 
to the square of the transverse size of the color dipole \cite{Low:1975sv,Nussinov:1975mw}. 
Note, that the average size of $c\bar c$ configurations involved in photoproduction 
of $J/\psi$ is 
significantly smaller than the $J/\psi$ size.
Such suppression of interactions 
of small dipoles is well known effect 
in  electrodynamics - for example,  a muonium can propagate through 
the media much easier than a positronium.

Within the leading $\ln Q^2$, $\ln (1/x)$  approximations of perturbative QCD  one expects (in difference from the Low - Nussinov model)
that the cross section of the 
interaction  of small dipoles 
with hadrons should 
increase rapidly with increase of invariant dipole-hadron energy $W_{\gamma N}=\sqrt{s}$ due to 
the growth of the gluon fields 
in hadron targets at small $x\propto s^{-1}$:
\beq
\sigma_{dip -  T}(x,d)={\pi^2\over 3}F^2 d^2\alpha_s (\lambda/d^2)xG_T(x,\lambda/d^2),
\label{dipole} 
\eeq
where $F^2= 3 (4/3)$ is the Casimir operator for the  two-gluon ($q\bar q$) dipole, 
and $\lambda \sim 4-9 $. 
For a dipole of a  
size $\sim$ 0.25  fm relevant for production of $J/\psi$ 
Eq.(\ref{dipole}) corresponds to energy dependence $\propto s^{0.2}$  and describes well behaviour
of the  both exclusive electro production of vector mesons and the inclusive  cross section of deep 
inelastic electron-proton scattering  scattering observed at HERA, for a review and references 
see \cite{Frankfurt:2005mc}. 

A naive extrapolation of the observed pattern to LHC energies
indicates that the strength of this interaction may reach values comparable
to that experienced by light hadrons, leading to a
new regime of strong interaction physics at the LHC
characterized by a strong absorption of small color dipoles by the media. On the other hand it is evident that to avoid conflict with probability conservation 
starting from some energies such
rapid increase of cross section should be tamed. 

So, the question is whether
it will be possible at LHC to observe this new perturbative QCD regime when 
the coupling constant is small
but the interaction is strong.
In practice, it is very difficult to devise
a high energy probe for virtualities of few GeV$^2$ for the hadron colliders 
especially  
for  the high energy strong interactions involving nuclei 
where gluon densities per unit area are higher and where 
new high gluon density physics should be enhanced.
Such a problem is absent for electron - ion colliders but these colliders are far in the future. 

An alternative which we  discuss here is to use ultraperipheral  collisions (UPC) of ions  
at the LHC in which one of the nuclei serves as a source of quasireal 
photons and another one as a target. The recently 
published study \cite{Baltz:2007kq} demonstrates that it is feasible to select UPC at the LHC 
and that the rates for many  processes of the dipole - nucleus interactions are high enough. 
This includes the process of  coherent photoproduction of \jp\cite{Frankfurt:2001db}. 
However, this process could only  be effectively studied up to relatively
 small energies $W_{\gamma N}\sim  {\rm \, 130 \,  GeV}$ due to the 
inability to  separate contributions due to the lower and higher energy photons 
emitted by two  colliding   ions. 
Here we suggest a strategy which avoids the above mentioned 
shortcoming of the coherent \jp production. It is based on the study  of the 
large momentum transfer 
$-t\equiv q^2\equiv (p_{\gamma} - p_{J/\psi})^2$ process:
$\gamma + A \to J/\psi + {\rm  \, gap}  + X$.
In addition to the theoretical advantages which we will explain below it also has some
appealing observational features. Observation of \jp and hadrons allows to determine 
unambiguously which  of the nuclei emitted the photon. As a result it is possible 
to observe the process up to  $W_{\gamma -N}\sim {\rm 1 TeV}$. Besides,  acceptance of the all three LHC  detectors 
which plan to study heavy ion collisions is sufficiently large for the discussed kinematics.
\begin{figure}[ht]
\centering\includegraphics[height=3cm,width=6cm]{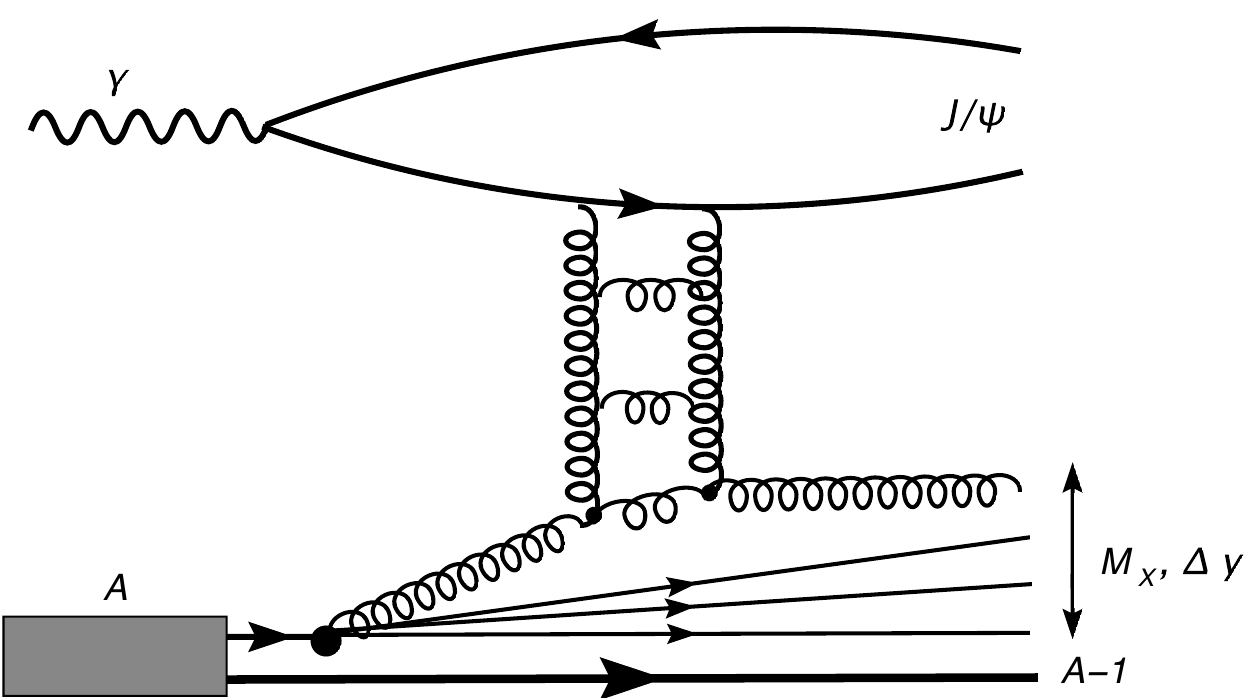}
\caption{The typical perturbative QCD diagram for  \jp production with gap.} 
\label{1}
\end{figure}

At first, let us briefly consider the main features of
the elementary reaction  $\gamma + p \to J/\psi + {\rm  \, gap}  + X$ in a case of a proton target
which was extensively studied at HERA. This process belongs to a class of reactions 
introduced in \cite{Frankfurt:1990nc,Mueller:1992pe}  where selection 
of large $q^2\gg 1/r_N^2$ ensures that transverse momenta 
remain large in all rungs of the gluon ladder  and therefore 
two gluons are attached  to one parton in the target, see Fig. \ref{1}. 
As a result the  cross section can be written in a factorized form as a product 
of the gluon density in the target 
(the kinematics can be chosen so as the correction due to scattering off 
the quarks which we will not write below explicitly will be small)  
and the elastic cross section 
of  scattering of a small dipole off a gluon:
\begin{eqnarray}
\frac {d\sigma_{\gamma T\to J/\psi X}} {dq^2 d\tilde{x}}=
\frac {d\sigma_{\gamma g \to J/\psi g}} {dq^2} g_{T}(\tilde{x},q^2),  \label{DGLAPBFKL}
\end{eqnarray}
where 
$g_{T}(\tilde{x},q^2)$ is the gluon distribution in the target, 
$\tilde{x}=q^2/(M_X^2 - m_N^2 +q^2)$ and $M_X$ 
is the invariant mass of the produced hadron system. Though it is not practical 
to measure $M_X^2$ directly, it can be expressed with a good accuracy 
through $\Delta y$, the rapidity interval occupied by the system $X$ as
\begin{eqnarray} 
\Delta y =\ln (\sqrt{q^2}/\tilde{x}m_N).
\label{gapdef}
\end{eqnarray}

 
The HERA data for the large $q^2$ and rapidity gap photoproduction of $J/\psi$ off the proton target
are consistent with the expectations of the models based on
the factorization
approximation in Eq.(\ref{DGLAPBFKL}) 
and QCD inspired  models for the $\gamma g \to J/\psi g$ amplitude, 
see Refs. \cite{Frankfurt:2008er}, \cite{Ivanov:2004ax} for recent data analyses and  references.

Now let us analyse the effects arising in a case of the nuclear target in which the 
color singlet dipole produced in the photon interaction with one of the nuclear nucleons 
passes through nuclear medium before to form the \jp. 
Firstly, we need to estimate the average dipole size, $d$,  which enters in the elementary  
$\gamma g\to gJ/\psi$ amplitude.
The overlapping integral between photon and charmonium wave functions 
is determined by c-quark momenta $k_t\sim m_c$.    For a two body $c\bar c$ system 
with Coulomb interaction one finds  
  $d \approx \sqrt{3/8}\pi /m_c$ for    $q^2\ll m_c^2$, which is similar to the estimate
  of $d \approx .25 \, {\rm fm}$ \cite{Frankfurt:1997fj}  for 
the 
process of forward \jp coherent photoproduction  with a charmonium realistic 
wave 
function.  
For large $q^2$ in the    Coulomb atom approximation we find 
$d^2(q^2)/d^2(0) \approx 4m_c^2/q^2$ for the dominant amplitude when 
gluons are attached to the same c-quark. Hence  we will use 
in our analysis the  interpolation formula
\beq
d^2(q^2)/d^2(0) \approx (1+q^2/4m_c^2)^{-1},
\label{size}
\eeq
 with $d_0= {\rm .25 fm }, m_c=1.5 \, {\rm GeV}$.

We choose the kinematics where
 $\tilde{x}\ge x_{sh}\equiv 0.01$, hence,  the shadowing and antishadowing effects in 
the nucleus parton density in the process of hard interaction 
are small, $g_{A}(\tilde x ,q^2)\approx Ag_{p}(\tilde x, q^2)$,
 and Eq.(\ref{DGLAPBFKL}) leads to 
a  cross section  which is approximately linear in A.

Now  we demonstrate that  diagrams 
where the ladder is attached to gluons belonging to different nucleons 
of the nucleus give a small contribution for realistic nuclei in spite of been enhanced by a combinatorial factor $\propto A^{1/3}$. For simplicity we consider  scattering at  central impact parameters for the rapidity interval given 
by Eq.(\ref{gapdef}). 
In this case one of the partons should carry  
$\beta_1$  light cone fraction of the nucleon momentum, which is  comparable to $ \tilde{x}$, 
while the second one is allowed to have any $\beta_2\ge \beta_1$. 
To obtain an upper limit for the ratio,   $R_2$, 
of contributions of these diagrams to that described by diagram in Fig.\ref{1} 
we assume that momentum 
transfers in the interactions with two partons of the nucleus are $q^2_1\sim q^2_2 \sim q^2/4$.  
Emission of two partons with 
$p_t=\sqrt{q^2}/2 $ 
to the same angular interval as for the single parton with $p_t=\sqrt{q^2}$  
requires  $\beta_1=\tilde{x}/2$ and $\beta_2\ge \beta_1$. 
Hence in this approximation $R_2$ is expressed through 
the number of gluons  in the cylinder of radius
 $d/2$ with $\beta_1 = \tilde{x}/2, \beta_2\ge \beta_1$ as (cf. Eq.(1)) :
\beq
R_2= {\pi [d(q^2)]^2\over 4} 
{g({\tilde{x}\over 2},{q^2\over 4})T(b=0)\over g(\tilde{x},q^2) }
\cdot 2\int_{{\tilde{x}\over 2}}^1 d\beta_2 g(\beta_2,{q^2\over 4}),
\eeq
where $T(b)=\int_{-\infty}^{\infty}\rho_A(b,z)dz$ is the standard thickness
function normalized to A.  A factor of two  in the nominator is due to the presence 
of two different attachments of  gluons to partons of the target.
For the \jp production,  using Eq.(\ref{size})
we find, for example, for $q^2=4 \rm{GeV}^2, \tilde{x}=x_{sh}$, 
$R_2 <  5\%$.
The factor $R_2$ stays small at larger $\tilde{x} $ due to decrease of parton densities with x.
The same conclusion is valid  for $\tilde{x} \ge 0.1$ where contribution of scattering off quarks 
becomes noticeable.
 Note also that the momentum balance in the case of interaction with two different partons is 
strongly different than in the case of interaction with a single parton leading to possibility 
to suppress this contribution using kinematic cuts.

Hence we conclude that, in the discussed kinematics,  significant deviations from the linear  A-dependence can arise only due to 
possible {\it inelastic} interactions of the dipole with other nucleons of the nucleus. The change of the probability of the gap survival with A:
\begin{eqnarray}
P^{gap}_A\equiv \frac {A_{eff}} {A}=\frac {d\sigma_{\gamma A\to J/\psi X}} {dq^2 d\tilde{x}}
/A\frac {d\sigma_{\gamma N\to J/\psi X}} {dq^2 d\tilde{x}},
\end{eqnarray}
which is formally a higher twist effect in $d^2$ has a physical meaning of the probability  for a small dipole to pass through the media without inelastic 
interactions for the energy $W_{\gamma N}$.

This effect can be expressed through the profile function for the 
dipole - nucleus scattering, $\Gamma_{dip, A} (x,d, {\vec b})$ ,
which is the Fourier conjugate of the elastic dipole - A amplitude. 
It is normalized so that $\sigma_{tot}(dip - A)(x,d) =2\int d^2b \Gamma_{dip, A}(x,d,{\vec b})$.  The range of  gluon $x$ probed in this case is of 
the order $x \equiv m_{J/\psi}^2/W_{\gamma N}^2$ (and somewhat smaller if one uses
 the charmonium model for the \jp wave function like in \cite{Frankfurt:1997fj}). 
In the dynamics driven by inelastic interactions 
$\left |\Gamma(x,d,{\vec  b})\right| \le 1$. 
Application of  S-channel unitarity (essentially the probability conservation, cf. \cite{Landau} ) allows one  to demonstrate that the  probability for the 
dipole not to interact inelastically is equal to $\left |1 -\Gamma(x,d,{\vec b})\right|^2$ leading to 
\beq
 A_{eff}= \int d^2b \, T({\vec b}) \left |1 -\Gamma(x,d,{\vec b})\right|^2.
\label{supfac}
\eeq
Here we neglect fluctuations in the size of the dipole which is a good approximation for the regime of moderate absorption where {\it average} interaction strength enters into the answer. In the case of large absorption the filtering effect takes place leading to enhancement of the contribution of small dipoles. A more detailed treatment will be given elsewhere.

Choice of kinematics with $\tilde{x}  > x_{sh} $ results in dominance of hard interaction
at small impact parameters. Thus, using a heavy nucleus as a target 
one can probe propagation of a small dipole through $\sim$ 10 fm of nuclear matter and determine $ \left |1 -\Gamma(x,d,{\vec b})\right|$.

To estimate the suppression effect as given by Eq.\ref{supfac}
we use two popular  complementary models for the interaction of a small size dipole with the matter.  One is the eikonal model where the small size dipole interacts via  multiple rescatterings off nucleons with the strength given by the dipole - nucleon total cross section. 
Deviations of the dipole - nucleus interaction from $\propto A$ is a higher twist effect since the interactions with  $n\ge 2$ - nucleons is $\propto d^{2n}$.  
Second  model  is the leading twist shadowing model which includes only   two gluon attachments to  the dipole.  In this case  
deviations from the  linear regime in $A$ are    due to soft interactions of the two gluons with the nucleus.

In the  eikonal model, 
neglecting fluctuations of the $c\bar c$ transverse size we obtain:
\beq
\Gamma(x,d,{\vec b})= 1 - \exp (-\sigma_{dip - N}(x,d) T({\vec b})/2),\eeq
where $\sigma_{dip - N}(x,d)$ is given by 
Eq.(\ref{dipole}). 
Since for heavy  nuclei 
$T(0) \approx  2 \, {\rm fm}^{-2}$,  $P^{gap}_A \approx \exp (-\sigma_{dip - N}(x,d) T(0))$
becomes small already for $\sigma \sim 5\,{\rm mb}$ which corresponds to $x \sim 10^{-3}$.
 Hence in this model a large suppression effect is expected which grows with $W_{\gamma N}$ 
and, for fixed $W_{\gamma N}$, decreases with increase of $q^2$, see Fig.2 (the curves
for $q^2=50 \, {\rm GeV}^2$ aim to illustrate the trend of the t-dependence of $P_A^{gap}$, the
actual measurement for this range of $t$ will require a long running time). 

An alternative model is the leading twist approximation over  parameter 
$\Lambda_{QCD}^2/(4m_c^2 + q^2)$ 
for the  dipole scattering off the nucleus which was used for the description of  
coherent $J/\psi$ production. Contrary to the eikonal approximation this approach
accounts for  essential nuclear modification 
of the nuclear parton distributions at small x.
Since in the leading twist Eq.(\ref{dipole}) describes 
inelastic dipole - nucleus cross section, 
the probability for a dipole of the size $d$ to pass through the nucleus without inelastic 
interactions
is 
\beq
P^{gap}_A={\frac {1} {A}} \int d^2b \, T({\vec b})
[ 1-\sigma_{dip - N}(x,d)
{g_A(x,Q^2,{\vec b})\over g_N(x,Q^2)}], 
\eeq
where $Q^2 =\lambda /d^2$ and $g_A(x,Q^2,{\vec b})$ is the gluon density of the nucleus 
in  impact parameter space 
($\int g_A(x,Q^2,{\vec b})d^2b=g_A(x,Q^2)$).
In the kinematic range $x\ge 3 \cdot 10^{-3}$ where 
shadowing effects 
 are still small one can 
unambiguously calculate the  shadowing correction as a function of ${\vec b}$
 through the diffractive gluon parton  distribution function (pdf), 
$g_{diff}(x,x_{\Pomeron},Q^2)$ which is measured in hard processes at HERA. 
Higher order rescatterings could be estimated by introducing 
$\sigma_{eff} (x,Q^2) = \int_x^{0.01} d x_{\Pomeron} g_{diff}(x,x_{\Pomeron},Q^2)/g_N(x,Q^2)$, 
for details see \cite{Frankfurt:2003zd}. 

\begin{figure}[ht]
\centering\includegraphics[height=5cm,width=6cm]{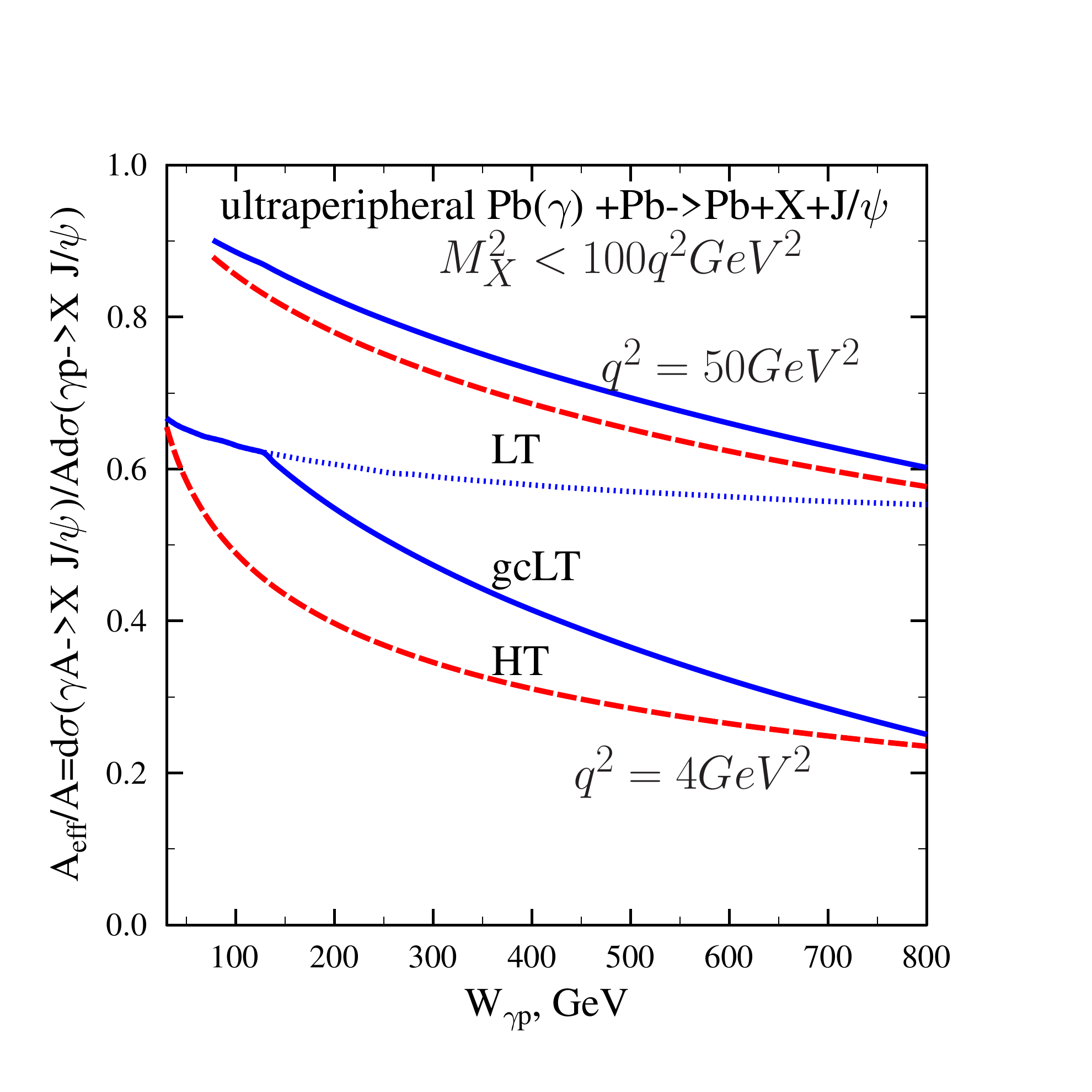}
  \caption{The rapidity gap survival probability as a function of 
$W_{\gamma N}$ for  $q^2=4 \, GeV^2$ and $q^2=50 \, GeV^2$}. 
 \label{gapsup}
 \end{figure}

The very small x and  low virtuality diffractive pdf's  one has to use for such an analysis 
are not reliable as they  involve extrapolations from larger $Q^2$ and $x$.
 A  straightforward application of the data leads to a very strong shadowing of $g_A$ 
and hence to a small absorption, see Fig.\ref{gapsup} (dotted line). 
However it is very difficult to envision leading twist (LT)  dynamics where partons of nucleons at a 
given impact parameter, $b$, would screen the nuclear
pdf below 
the maximal value of generalized gluon density $g_N(x,Q^2,\vec{\rho})$ of one nucleon at this $b$ (in the Glauber model  for the nucleon-deuteron interaction 
this condition corresponds to $\sigma (hD ) \ge \sigma (hN)$).  
For the limit of large A this implies a condition 
\begin{equation}
g_A(x,Q^2,{\vec b=0}) \ge g_N(x,Q^2, \vec{\rho}=0).
\label{limit}
\end{equation}
An effective way to implement this condition is to use the eikonal expression for
 screening of $g_A(x,Q^2,{\vec b})$:
 $$\frac {g_A(x,Q^2,{\vec b})} {g_N(x,Q^2)} = \frac {2} {\sigma_{eff}(x,Q^2)} 
[1-  \exp({-T({\vec b})\sigma_{eff}(x,Q^2)\over 2})],
$$ which 
leads to $g_A(x,Q^2,{\vec b})/g_N(x,Q^2) \le 2/\sigma_{eff}(x,Q^2)$ and find 
maximal value 
of $\sigma_{eff}(x,Q^2)$ which satisfies Eq.\ref{limit} for large $T(b)$. 
The $t$-dependence of the gluon generalized parton distribution $g_N(x,t)$ 
is measured at HERA in the exclusive vector meson production. For $x\sim 10^{-4}$  
exponential fits ($\exp(Bt)$) find $B\sim 5 \rm{GeV}^{-2}$.
Using this fit
we find $\sigma_{eff}^{max}=4\pi B\sim  \rm{27 \, mb}$. 
This geometrically constrained LT (gcLT) model leads to the 
solid curve in Fig.\ref{gapsup}.
One can see that the magnitude of the suppression predicted by such gcLT model is rather
close to that 
obtained
in the  eikonal model, the effect of absorption is large and strongly depends on energy and
$q^2$.

To determine for what kinematic range at LHC the measurements 
of cross section of the ultraperipheral process 
$A+A\to A+J/\psi +\,gap+X$
are feasible we used
the standard expression for 
the photon flux 
generated by the relativistic ions.
To calculate the cross section  of $\gamma p\to J/\psi +\,gap \,+X$ we adopted the QCD motivated
parametrization of \cite{Frankfurt:2008er} which fits all relevant HERA  
data using CTEQ6L parton distributions \cite{cteq6}.
$P_A^{gap}$ was calculated 
 within 
the models described above.
We imposed the
cut:  
$\tilde{x}_{min}\ge 0.01$ corresponding to 
$M_{X}\le 10\sqrt {q^2}$.

\begin{figure}[ht]
\centering\includegraphics[height=5cm,width=6cm]{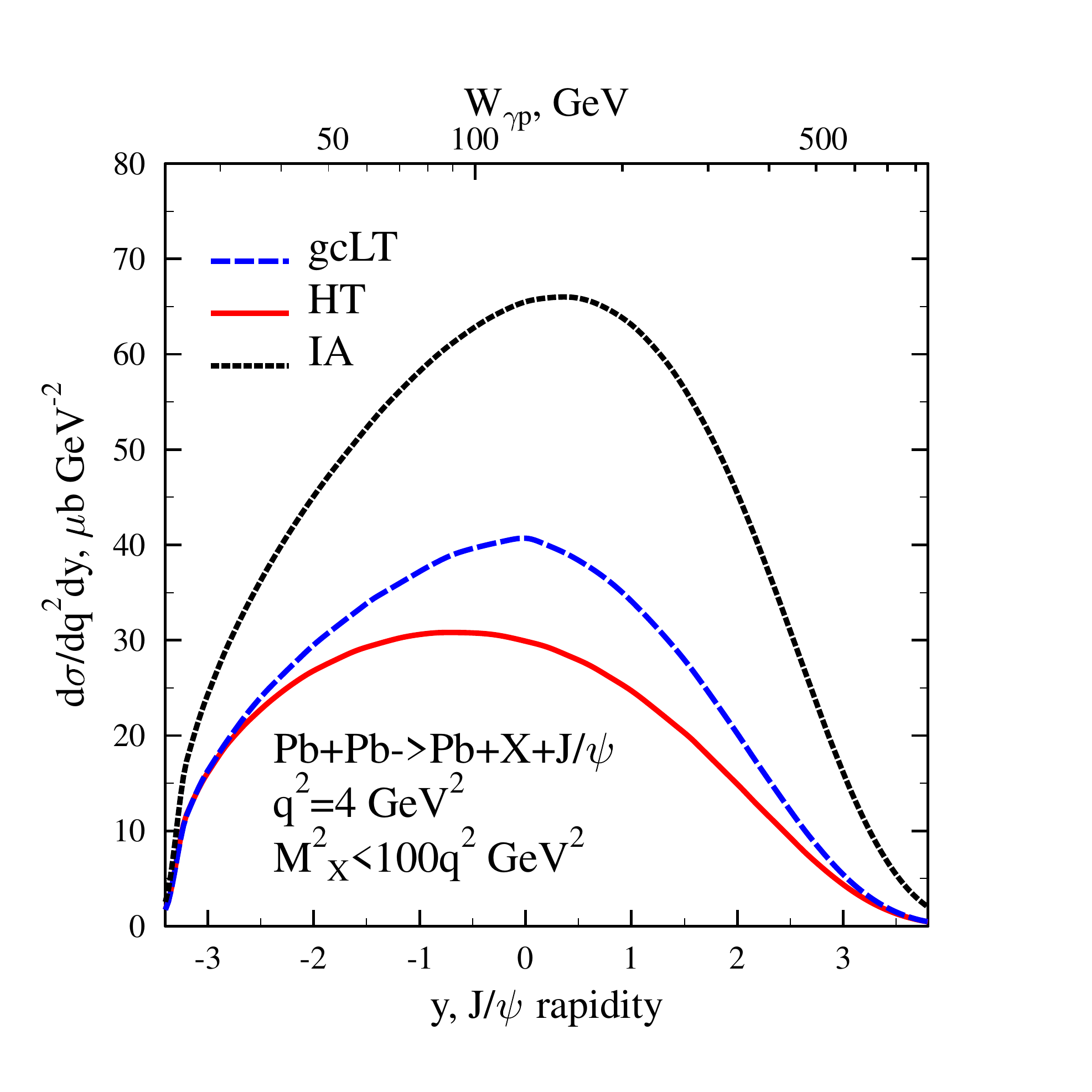}
 \caption{The rapidity distribution for the $J/\psi$ photoproduction in the
UPC $Pb+Pb\to Pb+J/\psi+gap+X$. }
 \label{cs}
\end{figure}

 Since it is possible to determine
experimentally which colliding ion serves as the source of photons we don't
account here for the symmetrical contribution from another ion which increases 
the counting rate by a factor of two. The results  presented in Fig.\ref{cs} indicate 
that for the planned integrated luminosity  for one month of running per 
year  $4.2\times 10^{5} \, {\rm mb}^{-1}$, it will be possible to measure $P_A^{gap}$ in 
a wide range of $q^2$ up to $W_{\gamma N} \sim$ 1 TeV.   
A much larger range 
of $q^2$ will be feasible for
 the process $\gamma+A \to \rho +\, {\rm gap} +X$ both due to a 
larger elementary cross section and absence of the suppression due 
to a small branching ratio for $J/\psi \to l^+l^-$. 
Measurements with  $\rho$'s will also allow to determine up to 
what $q^2$ the end point contribution which corresponds to very 
small $P_A^{gap}\propto  A^{-2/3}$ gives a noticeable contribution and at 
what $q^2$  the leading twist contribution leads to $P_A^{gap}(\rho)
 \approx P_A^{gap}(J/\psi)$.

In conclusion, we find that a  study of the \jp in UPC with rapidity gaps will allow to determine 
the pattern of absorption for the interaction of small dipoles with $0.15 fm\le 
d \le 0.25\, {\rm  fm}$ up to $W_{\gamma N}\sim 1 \rm{TeV}$ which corresponds to interaction with gluon fields 
with $x$ down to $10^{-6}$ for virtualities of $\sim 4 \div 10 \,{\rm GeV}^2$. The
ability to vary both $q^2$ and $W_{\gamma N}$ will allow to separate the  regions where 
opacity should be 
large and where the high energy color transparency will hold. Extension of these measurements 
to smaller $q^2$ will help to get better understanding of soft diffractive dissociation induced 
by a small dipole.
The discussed \jp measurements will also serve as an important benchmark for the studies 
of \jp production in the heavy ion collisions at the LHC
 as well help to understand the possible role of small dipoles in generating strongly fluctuating 
pedestal in the inelastic $pp$ collisions.

 We would like to thank V.~Guzey, A.~Mueller and T.~Rogers for useful discussions.
This work was supported in part by the US DOE Contract Number
DE-FG02-93ER40771 and BSF; 
 M. Zhalov research was also supported by CERN-INTAS grant no 05-103-7484.

\end{document}